\documentclass[aps,prl,preprintnumbers,showpacs,twocolumn]{revtex4}

\usepackage{amsmath}
\usepackage{amssymb}
\usepackage{bm}
\usepackage{dcolumn}
\usepackage{epsf}
\usepackage{mathrsfs}
\usepackage{verbatim}
\usepackage[english]{babel}

\begin{document}

\title{Spin relaxation in the impurity band of a semiconductor in the external magnetic field}
\author{I.S.\;Lyubinskiy}
\email{ilyalyu@mail.ru}
\affiliation{A.F.\;Ioffe Physical Technical Institute, 194021 St.\;Petersburg, Russia}
\date{September 1, 2007}

\begin{abstract}
Spin relaxation in the impurity band of a 2D semiconductor with spin-split spectrum in the ex\-ter\-nal magnetic field is con\-sid\-ered. Several mechanisms of spin relaxation are shown to be relevant. The first one is attributed to phonon-as\-sisted transitions between Zeeman sublevels of the ground state of an isolated impurity, while other mechanisms can be described in terms of spin precession in a ran\-dom mag\-netic field during the electron mo\-tion over the impurity band. In the later case there are two contributions to the spin relaxation: the one given by optimal impurity configurations with the hop-waiting time inversely proportional to the external magnetic field and another one re\-lat\-ed to the electron motion on a large scale. The average spin re\-laxa\-tion rate is cal\-culated.
\end{abstract}


\maketitle

%% --------------------------------------------------------
%% Introduction
%% --------------------------------------------------------

Spin dynamics in semiconductors has attracted much attention in the last decades \cite{rev1,rev2}. In particular, a number of experimental \cite{awsch1,awsch2,dzh1,dzh2,dzh3} and theoretical \cite{kavokin,shkl,our,tambo} works are devoted to the investigation of spin relaxation in the impurity band of a semiconductor. An increasing interest to this problem is motivated by experimental observation of up-to-300ns spin lifetimes in n-doped bulk GaAs and GaAs/AlGaAs heterostructures \cite{awsch1,awsch2,dzh1,dzh2}, which makes them good candidates for the use in possible spintronics applications. Yet, a consistent theory of spin relaxation in the impurity band is still to be developed. Depending on the donor concentration, spin relaxation in the impurity band might be driven either by hyperfine interaction or spin-or\-bit coupling. Since the nuclear spin relaxation time is typically very long, hyperfine interaction can be treated as a ran\-dom-in-space static magnetic field with the associated spin precession frequency $\omega_N \sim A/\sqrt{N}$, where $A$ is the hyperfine coupling constant and $N$ is the number of nuclei within the volume occupied by the wave function \cite{Bhyp} (the directions of the random magnetic field for electrons located on different impurities are not correlated). In the case of spin-or\-bit coupling, the associated spin precession frequency $\boldsymbol{\omega}_{\mathbf p}$ is a power function of the electron momentum $\mathbf p$ \cite{rashba,dress,dressqwell} (in the 2D case, $\boldsymbol{\omega}_{\mathbf p}$ is linear in $\mathbf p$). As a result, spin-orbit coupling leads to spin rotation in the process of phonon-as\-sist\-ed hops from one impurity to another by the angle $\boldsymbol{\phi} \approx \boldsymbol{\omega}_{\mathbf p_0} \Delta r/v_0$, where $\Delta r$ is the distance between impurities and $\mathbf p_0 = m \mathbf v_0$ is the under-the-barrier momentum. There are several mechanisms of spin relaxation in the impurity band. As in quantum dots, spin relaxation might be driven by phonon-as\-sist\-ed transitions between Zeeman sublevels  of the ground state of separate impurities. Other mechanisms, that involve electron hops from one donor to another, are specific for the impurity band. In Ref.\;\cite{dzh3}, the spin relaxation rate was estimated as:
\begin{align}
1/\tau_S \sim \omega_N^2\tau_{hc},~~1/\tau_S \sim \phi^2/\tau_{hc} \label{tauSangdif}
\end{align}
for the case of hyperfine interaction and spin-orbit coupling respectively and the characteristic hop waiting time $\tau_{hc}$ was assumed to depend only on the average distance between impurities. These equations are based on the classical picture of the angular spin diffusion in a random magnetic field (in the case of hyperfine interaction, the direction of spin precession changes randomly after each hop; in the case of spin-orbit coupling, the spin rotates by a certain angle in a random direction in the process of a hop). However, this approach does not account for the exponential variation of the hop waiting times:
\begin{align}
\tau_{h1} &= \tau_0 \exp\left(2\Delta r/a\right), \label{t1}
\\
\tau_{h2} &= \tau_0 \exp\left(2\Delta r/a+\Delta \mathcal E/T\right) \label{t2}
\end{align}
for phonon emission and absorbtion respectively (here $\Delta r$ is the distance between impurities, $\Delta \mathcal E$ is the distance between the energy levels, $a = \epsilon\hbar^2/2me^2$ is the Bohr radius, and $T$ is the temperature). The main consequence of such inhomogeneity is that it is impossible to introduce an universal time scale for the system under consideration. This fact is confirmed by about ten-fold decrease of the experimentally measured spin correlation time in the bulk GaAs at the crossover from hyper\-fine-in\-ter\-action-in\-duc\-ed to spin-orbit-induced spin relaxation (see Fig.\;3 in Ref.\;\cite{dzh3}). The effects of the inhomogeneity on the spin dynamics in the absence of the external magnetic field were considered in Refs.\;\cite{shkl,our} for the system with spin-split spectrum. In particular, it was found that there are two essentially different contributions to the spin relaxation: the one related to electron hops over the pairs of impurities with the size of the order of the Bohr radius and another one related to the motion over a long distance.

In this letter, we calculate the average spin relaxation rate for the mechanisms discussed above in the presence of the external magnetic field. The use of the averaged relaxation rate is justified if the relaxation is slow enough so that an electron can walk over a large distance during the spin relaxation time $\tau_S$ (in the opposite case the spin relaxation is governed by escape from the regions with slow relaxation to the regions with fast relaxation \cite{our}). The corresponding condition is \cite{our} $\tau_S \gg \tau_C$ (here $\tau_C = \tau_0 \exp\left(C\xi_0\right)$ is the hop waiting time for so-call\-ed critical bond, $\xi_0 = \sqrt[3]{4 L_d^2W/a^2T}$, $C$ is the dimensionless coefficient \cite{shklbook}, $W = e^2/\epsilon L_d$ is the characteristic width of the impurity band, and $L_d=n_d^{-1/2}$ is the average distance between impurities). We assume that spin precession in the external magnetic fields is sufficiently fast $\Omega_0 \tau_S \gg 1$ (here $\Omega_0$ is the spin precession frequency in the external magnetic field $B$). In this case, the components of the spin perpendicular to the magnetic field are suppressed due to fast precession, and hereafter they will be neglected. We assume that the temperature is sufficiently small $T \ll W$, so that we can neglect activation to the conduction band. We also assume that $\hbar \Omega_0 \ll T$, neglect elec\-tron-elec\-tron interaction, and treat donors as 2D coulomb centers.

Our point is that over a wide range of magnetic fields the relevant time scale $\tau_{hc}$ for the problem under consideration is given by:
\begin{equation}
\tau_{hc} = 1/\Omega_0. \label{tauC}
\end{equation}
Indeed, a common feature of the relaxation mechanisms based on the angular spin diffusion in a random magnetic field is that they are suppressed by applying a longitudinal magnetic field with the associated spin precession frequency larger than the inverse correlation time of the random magnetic field. In the simplest case of a pair of impurities with the hop waiting times $\tau_{h1} = \tau_{h2} = \tau_h$ ($\Delta \mathcal E \ll T$), the spin relaxation rate is proportional to $\Delta \Omega^2 \tau_h/\left(1+\Omega_0^2 \tau_h^2\right)$, where $\Delta \Omega$ is the spin precession frequency in the random magnetic field (in the case of hyperfine interaction $\Delta \Omega \approx \omega_N$; in the case of spin-or\-bit coupling $\Delta \Omega \approx \Omega_0 \phi$, as shown below). The contribution of the pairs to the spin relaxation increases exponentially with $\Delta r$ for $\tau_h < 1/\Omega_0$ and decrease for $\tau_h > 1/\Omega_0$. Taking into account Eqs.\;\eqref{tauSangdif} and \eqref{tauC}, we can estimate the spin relaxation rate on the pairs of impurities as:
\begin{equation}
1/\tau_S \sim \nu \omega_N^2/\Omega_0,~~1/\tau_S \sim \nu \phi^2 \Omega_0, \label{estimates}
\end{equation}
 for the case of hyperfine interaction and spin-orbit coupling respectively (here $\nu \sim \left(a/L_d\right)^2 T/W$ is the share of the optimal pairs). We also consider other relevant mechanisms of spin relaxation that are related to the electron motion over a long distance and spin-flip processes.

%% --------------------------------------------------------
%% Hamiltonian
%% --------------------------------------------------------

Let us proceed to the rigorous formulation of the problem. First, we consider the system with spin-split spectrum. The Hamiltonian of the system is
\begin{equation}
\hat H = \hat H_0+\hat H_{\rm  ph}+\hat H_{\rm e-ph}, \label{sum}
\end{equation}
where $\hat H_0$ and $\hat H_{\rm  ph}$ are the Hamiltonians of an electron and phonons respectively, and the last term on the right-hand side describes electron-phonon interaction. The Ha\-miltonian of an electron is given by:
\begin{equation}
\hat H_0 = \frac{\mathbf p^2}{2m}+U\left(\mathbf r\right)+ \hbar \boldsymbol{\sigma}\boldsymbol{\Omega}_0/2+ \hbar \boldsymbol{\sigma}\hat\alpha\mathbf p/2mL_S, \label{H0}
\end{equation}
where $U\left(\mathbf r\right)$ is the impurity potential, $L_S$ is the length characterizing the strength of the spin-orbit coupling, $\hat\alpha$ is the dimensionless tensor with the components of the order of unity, and $\boldsymbol{\sigma}$ is the vector of Pauli matrices. The last term on the right-hand side is a combination of the Bychkov-Rashba spin-orbit coupling \cite{rashba} and Dresselhaus spin-orbit coupling averaged over the electron motion in the direction perpendicular to the quantum well \cite{dress,dressqwell}. The Hamiltonians of phonons and electron-phonon interaction are given by:
\begin{align}
\hat H_{\rm ph} &= \sum_{\mathbf q} \hbar s q \hat b_\mathbf q^+ \hat b_\mathbf q,
\\
\hat H_{\rm e-ph} &= \sum_{\mathbf q} C_n \sqrt{q^n/V}\left[ e^{i\mathbf q\mathbf r} \hat b_\mathbf q + e^{-i\mathbf q\mathbf r} \hat b_\mathbf q^+ \right],
\end{align}
where $\hat b_\mathbf q^+$ and $\hat b_\mathbf q$ are phonon creation and annihilation operators, $\mathbf{q}$ is the phonon wave vector, $s$ is the sound velocity, $C_n$ is the coefficient characterizing the strength of the electron-phonon interaction, $V$ is the system volume, and $n = \pm 1$ for deformation and piezoelectric phonons respectively. For the following consideration it is convenient to make a transformation, which cancels spin-orbit coupling to the first order in parameters $1/L_S$ and $\Omega_0$:
\begin{equation}
\hat H' = e^{i\boldsymbol{\sigma}\hat\alpha\mathbf r/2L_S} \hat H e^{-i\boldsymbol{\sigma}\hat\alpha\mathbf r/2L_S}. \label{convert}
\end{equation}
As a result,
\begin{equation}
\hat H_0' = \frac{\mathbf p^2}{2m}+U\left(\mathbf r\right)+ \hbar \boldsymbol{\Omega}_0 \boldsymbol{\sigma} /2 + \hbar  \left[\boldsymbol{\Omega}_0 \times \hat\alpha \mathbf r/L_S \right]\boldsymbol{\sigma}/2, \label{H0I}
\end{equation}
while $\hat H_{\rm  ph}$ and $\hat H_{\rm e-ph}$ are not modified.

%% --------------------------------------------------------
%% 1 impurity
%% --------------------------------------------------------

First, let us consider the spin relaxation caused by pho\-non-as\-sist\-ed transitions between Zeeman sublevels of the ground state of an isolated impurity. In this case the spin relaxation rate coincides with the transition rate obtained by the Fermi golden rule:
\begin{equation}
\frac{1}{\tau_S} = \frac{2\pi}{\hbar}\int N_{\mathbf q} W_{\mathbf q}\,\delta\left(\hbar\Omega_0-\hbar s q\right) {Vd\mathbf q}/{\left(2\pi\right)^3},
\end{equation}
where $W_{\mathbf q} = \left| \left<\Psi_+\left|\exp\left(i\mathbf q \mathbf r\right) \right|\Psi_-\right>\right|^2$, $N_{\mathbf q}$ is the phonon occupation number, $\Psi_{\pm} = \Psi_{0\pm}+\delta\Psi_{\pm}$,
\begin{equation}
\Psi_{0\pm} = \sqrt{\frac 2 {\pi a^2}} \: e^{-r/a}  \left|\pm \right> \label{Psi0}
\end{equation}
are the eigenfunctions of an electron at a Coulomb center,
\begin{equation}
\delta\Psi_{\pm} = \left(\mathcal E_0  - \hat H_C\right)^{-1}\left[\hat\alpha \mathbf r/L_S \times \hbar\boldsymbol{\Omega}_0/2 \right] \boldsymbol{\sigma} \Psi_{0\pm} \label{Psi1}
\end{equation}
are the corrections due to the last term on the right-hand side in Eq.\;\eqref{H0I}, $\left|\pm\right>$ are the spinors, $\hat H_C = \mathbf p^2/2m-e^2/\epsilon r$ is the Hamiltonian of an electron at a Coulomb center, and $\mathcal E_0=\hbar^2/2ma^2$ is the binding energy. In the case of small magnetic fields $\Omega_0 a \ll s$ and $\hbar\Omega_0 \ll T$, we get:
\begin{align}
\frac{1}{\tau_S} =& \frac{T}{\hbar} \left(\frac{a}{L_S} \frac{a C_n }{ \mathcal E_0} \right)^2 \left(\frac{\Omega_0}{s} \right)^{5+n} {g\left(\mathbf e_0\right) I_0^2}/{8}, \label{vspinflip}
\\
g\left(\mathbf e_0\right) =& \sum_{i=x,y}\left|\mathbf e_0 \times \hat\alpha \mathbf e_{i}\right|^2,
\end{align}
where $I_0 = \left< \Psi_0\right|\mathbf r \left(\mathcal E_0+\hat H_C\right)^{-1} \mathbf r \left|\Psi_0 \right>\mathcal E_0/a^2 \sim 1$ and $\mathbf e_0$ is the unit vector in the direction of the external magnetic field. This dependence of the spin relaxation rate on the external magnetic field is in accordance with the results obtained in Ref.\;\cite{kha_qd} for the circular quantum dot with piezoelectric phonons ($n = 1$).

%% --------------------------------------------------------
%% 2 impurities
%% --------------------------------------------------------

Next, let us consider the spin relaxation on a pair of impurities caused by spin precession in the random magnetic field. From the Hamiltonian \eqref{H0I} one can derive an equation, describing spin dynamics:
\begin{equation}
\partial\mathbf S/\partial t = \left[\left(\boldsymbol{\Omega}_0+\Delta\boldsymbol{\Omega}\left(t\right)\right) \times \mathbf S\right], \label{dsdt}
\end{equation}
where $\Delta\boldsymbol{\Omega}\left(t\right) = \left[\boldsymbol{\Omega}_0 \times \hat\alpha \mathbf r\left(t\right)/L_S \right]$ and the position of the electron $\mathbf r\left(t\right)$ takes two value: $\mathbf r_1$ or $\mathbf r_2$ (here $\mathbf r_{1,2}$ are the positions of the impurities). To find the random magnetic field correlator
$\kappa\left(t\right) = \left\langle \Delta \boldsymbol{\Omega} \left(t\right) \Delta \boldsymbol{\Omega} \left(0\right) \right\rangle$, we use the kinetic equation for an electron on a pair of impurities:
\begin{equation}
dn_1/dt = -dn_2/dt = n_2/\tau_{h2}-n_1/\tau_{h1}, \label{kin}
\end{equation}
where $n_{1,2}$ are the probabilities to find an electron at im\-purity $1$ and $2$ respectively. Using the Green function of the kinetic equation, we get:
\begin{equation}
\kappa\left(t\right) = \frac{ \Delta \Omega^2 }{4\cosh^2\left(\Delta \mathcal E/2T\right)} \exp\left(-t/\tau_{h}\right), \label{corr}
\end{equation}
where $\Delta \boldsymbol{\Omega} = \left[\boldsymbol{\Omega}_0 \times \hat\alpha \Delta \mathbf r/L_S \right]$, $\Delta \mathbf r = \mathbf r_1-\mathbf r_2$, and $1/\tau_h = 1/\tau_{h1}+1/\tau_{h2}$. Treating the term proportional to $\Delta\boldsymbol{\Omega}\left(t\right)$ in Eq.\;\eqref{dsdt} as a perturbation and using Eq.\;\eqref{corr}, we get the following evolution equation for the component of the spin parallel to the external magnetic field:
\begin{equation}
{\partial S_{\|}}/{\partial t} = - \int \kappa\left(t'\right) \cos\left(\Omega_0 t'\right) S_{\|}\left(t-t'\right) dt',
\end{equation}
The spin relaxation rate on a pair of impurities is
\begin{equation}
1/\tau_S\left(\Delta \mathbf r, \Delta\mathcal E\right) =
\frac{ \Delta \Omega^2 }{4\cosh^2\left(\Delta \mathcal E/2T\right)} \frac{\tau_{h}}{1+\Omega_0^2\tau_{h}^2}. \label{tauSbell}
\end{equation}

%% --------------------------------------------------------
%% SO - intermediate fields
%% --------------------------------------------------------

Depending on the strength of the external magnetic field, several regimes of spin relaxation can be realized. In the case $\Omega_0 < 1/\tau_0$, the main contribution to the spin relaxation rate comes from the pairs of impurities with $\left|\Delta \mathcal E \right| \le T$ and $\tau_h \approx 1/\Omega_0$ (as follows from Eq.\;\eqref{tauSbell}, this contribution is proportional to the first power of the external magnetic field). The average spin relaxation rate is
\begin{equation}
1/\tau_S = \int 1/\tau_S\left(\Delta \mathbf r, \Delta\mathcal E\right) d\Delta\mathbf r d \Delta \mathcal E / W L_d^2. \label{tauSave}
\end{equation}
Substituting Eq.\;\eqref{tauSbell} into Eq.\;\eqref{tauSave}, we get:
\begin{equation}
\frac{1}{\tau_S} = \frac{\pi^2}{64} \Omega_0 \frac TW
\left(\frac a {L_S} \frac a{L_d}\right)^2 \ln^3\left(\frac 1{\Omega_0\tau_0}\right) g\left(\mathbf e_0\right). \label{tauSaveSmall}
\end{equation}

%% --------------------------------------------------------
%% SO - high fields
%% --------------------------------------------------------

In the case $\Omega_0 > 1/\tau_0$, we can neglect the unity in the denominator of Eq.\;\eqref{tauSbell}. As a result,
\begin{equation}
1/\tau_S\left(\Delta \mathbf r, \Delta\mathcal E\right) =
\frac 1{\tau_0}  \frac{\left[\mathbf e_0 \times \hat\alpha \Delta \mathbf r/L_S \right]^2 \exp\left(-2\Delta r/a\right)}{1+\exp\left(\Delta \mathcal E/T\right)}. \label{tauSLargeO}
\end{equation}
In this case the main contribution to the spin relaxation rate comes from the pairs of impurities with $\Delta r \le a$ and $\left|\Delta \mathcal E \right| \le T$. Substituting Eq.\;\eqref{tauSLargeO} into Eq.\;\eqref{tauSave}, we get:
\begin{equation}
\frac 1 {\tau_S} = \frac{3\pi \ln 2}{4}\frac{1}{\tau_0} \frac TW \left(\frac a {L_S} \frac{a}{L_d}\right)^2 g\left(\mathbf e_0\right). \label{tauSaveLarge}
\end{equation}
Thus, the spin relaxation rate is saturated at large $\Omega_0$.

%% --------------------------------------------------------
%% SO - orbital motion
%% --------------------------------------------------------

In deriving Eq.\;\eqref{tauSaveSmall}, we assumed that optimal pairs are separated from the rest of the system, i.e. that an electron makes many hops over the pair before it leaves it. For this assumption to be valid, it is required that $\Omega_0\tau_C \gg 1$. In the opposite case, an electron motion on a large scale gives the leading contribution to the spin relaxation rate. This contribution is proportional to the electron diffusion coefficient \cite{shkl,our}:
\begin{equation}
1/\tau_S \sim D \sim 1/\tau_C.
\end{equation}
The influence of the external magnetic field on the diffusion coefficient is well known \cite{shklbook}. At low fields $R_C \gg a\xi_0$ (here $R_C$ is the cyclotron radius and $a\xi_0$ is the optimal hopping length \cite{shklbook}), it can be described in terms of bending of the tunnelling electron trajectory by the external magnetic field perpendicular to the quantum well, which effectively increases the distance between impurities:
\begin{equation}
a \xi_0 \to a \xi_0' \left(H\right)= a \xi_0\left[1+\frac{1}{60}\left(\frac{a\xi_0}{R_C}\right)^2 \right]. 
\end{equation}
As a result,
\begin{equation}
1/\tau_S\left(H\right) = \left(1/\tau_S\right) \exp\left[-\frac{1}{60} \left(\frac{a}{R_C}\right)^2 \xi_0^3 \right].
\label{tauSLargeScale}
\end{equation}

%% --------------------------------------------------------
%% HI - 2 impurities
%% --------------------------------------------------------

Next, let us briefly consider the spin relaxation caused by hyperfine interaction. In this case, Eq.\;\eqref{tauSbell} still can be used with the replacement $\Delta \Omega \to \Delta \Omega' = 2^{1/2} \omega_N$. Using Eq.\;\eqref{tauSbell} and following the same procedure as before, we get:
\begin{align}
\frac{1}{\tau_S} &= \frac{\pi^2}{8}\frac{A^2}{N\Omega_0} \frac TW \left(\frac a{L_d}\right)^2 \ln\left(\frac 1{\Omega_0\tau_0}\right),
\\
\frac{1}{\tau_S} &= \pi\ln 2      \frac{A^2}{N\Omega_0^2\tau_0} \frac TW \left(\frac{a}{L_d}\right)^2
\end{align}
for the case $\Omega_0 < 1/\tau_0$ and $\Omega_0 > 1/\tau_0$ respectively.

%% --------------------------------------------------------
%% HI - orbital motion
%% --------------------------------------------------------

In the case of small magnetic fields $\Omega_0\tau_C \ll 1$ the spin relaxation rate can be estimated as
\begin{equation}
1/\tau_S = \int A^2t/N dP\left(t \right), \label{tauS1imp}
\end{equation}
where $P\left( t \right)$ is the probability for an electron to spend time $t$ at an impurity. This probability can be replaced with the probability that an impurity is separated from the rest of the system by the length $a\ln \left(t/\tau_0\right)$ in the coordinate space and $T \ln \left(t/\tau_0\right)$ in the energy space. Assuming that the form of the surrounding empty area is given by $\left|\Delta r/a+\Delta \mathcal E/T \right| \le \ln\left[t/\tau_0\right]$, we get:
\begin{equation}
P\left(t\right) = \exp\left(-\ln^3\left[t/\tau_0\right] 8\pi/3\xi_0^3\right).
\label{prob}
\end{equation}
Substituting Eq.\;\eqref{prob} into Eq.\;\eqref{tauS1imp} and using saddle-point approximation, we get:
\begin{equation}
1/\tau_S \approx \left(\pi/32\right)^{1/4} \left(A^2/N\right) \xi_0^{3/4} \exp\left(\frac 23 \sqrt{\xi_0^{3}/8\pi}\right).
\end{equation}
The influence of the external magnetic field on the orbital motion can be accounted for by the replacement:
\begin{equation}
a \xi_0 \to a \xi_0' \left(H\right)= a \xi_0\left[1+\frac{1}{60}\left(a\sqrt{\xi_0^{3}/8\pi}/R_C\right)^2 \right].
\end{equation}
Here $a\sqrt{\xi_0^{3}/8\pi}$ is the size in space of the empty area at the saddle point. As a result,
\begin{equation}
1/\tau_S\left(H\right) = \left(1/\tau_S\right) \exp\left[\frac 1{60}\left(\frac{a}{R_C}\right)^2 \left(\xi_0^{3}/8\pi\right)^{3/2}\right]. \label{tauSLargeScale}
\end{equation}

%% --------------------------------------------------------
%% Conclusion
%% --------------------------------------------------------

To conclude, the theory of spin relaxation in the impurity band of a 2D semiconductor in the external magnetic field is presented. It is shown that spin precession in the external magnetic field enhances spin-or\-bit-induced and suppresses hyperfine-in\-teraction-induced spin relaxation. For spin orbit coupling, the relaxation rate is linear in $B$ over a wide range of parameters, while the dependence on the direction of the external magnetic field is the same for all spin-or\-bit-in\-duc\-ed relaxation mechanisms. For hyperfine interaction, the spin relaxation rate is inversely proportional to the external magnetic field.

The author is grateful to A.P.\;Dmitriev and V.Yu.\;Ka\-cho\-rov\-skii for useful critiques. This work has been supported by RFBR, by grants of RAS, and by a grant of the Russian Scientific School.

%% --------------------------------------------------------
%% Bibliography
%% --------------------------------------------------------

\end{document}